\documentclass{article}
\usepackage{amssymb}
\usepackage{bm}
\usepackage[dvips]{graphicx}
\begin{document}

\title{Parametric instability of a magnetic junction under modulated spin-polarized current}

\author{E. M. Epshtein\thanks{E-mail: epshtein36@mail.ru}, P. E. Zilberman\\ \\
V. A. Kotelnikov Institute of Radio Engineering and Electronics\\
of the Russian Academy of Sciences, Fryazino, 141190, Russia}

\date{}

\maketitle

\abstract{The stability is analyzed of the magnetic junction collinear
configurations against small fluctuations under amplitude-modulated
current with CPP mode. High spin injection is assumed. Under parametric
resonance conditions, with the modulation frequency twice the precession
frequency, instability is possible of one, or another, or both the
collinear configurations. When the dc component of the current density
exceeds the instability threshold of the antiparallel configuration, the
parametric instability is suppressed by nonparametric one which is induced
by the dc current. The parametric instability manifests itself as lowering
the threshold of the dc current density in presence of the high-frequency
current, such an effect has been observed in experiments repeatedly.}\\
\\

Studying the magnetic junctions under spin-polarized current showed a
number of interesting phenomena, such as the switching of magnetic
configuration~\cite{Katine}, spin-wave generation~\cite{Tsoi}, etc. Such
effects may appear in nanoscale level, because the corresponding
characteristic lengths, namely, the exchange length and the spin diffusion
length, are of order of 10 nm~\cite{Bass}. This allows using the effect
for high-density information recording by electric current, which is
unattainable for switching magnetic elements by magnetic field alone.

The current-driven switching magnetic junctions is accompanied often with
the magnetization oscillation and the other high-frequency effects (see,
e.g.,~\cite{Tsoi,Krivorotov,Ralph,Xiao}). In this connection, an
interesting problem occurs, namely, the influence of spin-polarized
high-frequency current on magnetic junctions.

The microwave effect on the switching magnetic structures by
spin-polarized current was studied in
Refs.~\cite{Rivkin,Cui,Florez1,Florez2,Biziere}. Lowering threshold value
of dc current density was found in presence of a high-frequency current.
Nowadays, interpretation of this effect is restricted mainly with
qualitative arguments and numerical simulations without developing any
consistent theory. Without having pretensions to constructing a
comprehensive theory, we should like to pay attention to a possible
mechanism of the effect related with occurring instability of the magnetic
junction configuration under parametric resonance conditions.

Let us consider a magnetic junction consisting of a pinned ferromagnetic
layer 1, free ferromagnetic layer 2, and nonmagnetic layer 3 closing the
electric circuit. Amplitude-modulated electric current with density
\begin{equation}\label{1}
  j(t)=\overline j+\widetilde j\cos\Omega t
\end{equation}
flows perpendicular to the layers.

We assume $\widetilde j\leq\overline j$, so that the total current flows always in
the same direction, corresponding to the conduction electron flux in the
$1\to2\to3$ direction. We take, also, $\Omega\tau\ll1$, $\tau$ being the
electron spin relaxation time, that allows describing the alternative
current effect by the same equations as the direct current one with
changing constant current density by time-dependent one, $j(t)$.

\begin{figure}
\includegraphics{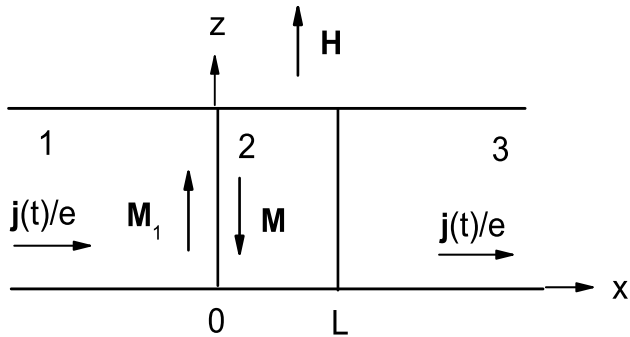}
\caption{Scheme of the magnetic junction.}\label{fig1}
\end{figure}

We take a coordinate system with $x$ axis perpendicular to the layers,
$yz$ plane parallel to layers, and $x$ coordinate counting off the
interface between 1 and 2 layers (Fig.~\ref{fig1}). The pinned layer
magnetization vector $\mathbf{M}_1$ is directed along $z$ axis. The free
layer magnetization vector $\mathbf M$ originally (without current) is
collinear (parallel or antiparallel) to the pinned layer magnetization
vector $\mathbf{M}_1=\{0,\,0,\,M_1\}$. We are to investigate stability of
such a state in presence of the current.

The free layer thickness $L$ is assumed to be small compared to the spin
diffusion length in that layer and the domain wall thickness. In such a
case, the macrospin approximation is valid, in which the free layer
magnetization is taken as spatially uniform, while the presence of the
current is taken into account by means of additional terms proportional to
the current in the equations of the motion for the lattice magnetization
(see Refs.~\cite{Gulyaev1,Gulyaev2} for detail).

In the linear approximation on small deviations of the $\mathbf M$ vector
from the collinear position, the Landau-Lifshitz equations take the form
\begin{eqnarray}\label{2}
  &&\frac{d\hat{M}_x}{dt}+\kappa\overline{\hat{M}}_z\frac{d\hat{M}_y}{dt}
  +\gamma\left(H_z+H_a\overline{\hat{M}}_z\right)\hat{M}_y\nonumber \\
  &&+P(t)\hat{M}_y+K(t)\overline{\hat{M}}_z\hat{M}_x=0,
\end{eqnarray}
\begin{eqnarray}\label{3}
  &&\frac{d\hat{M}_y}{dt}-\kappa\overline{\hat{M}}_z\frac{d\hat{
  M}_x}{dt}-\gamma\left(H_z+H_a\overline{\hat{M}}_z+4\pi M\overline{\hat{M}}_z \right)
  \hat{M}_x\nonumber \\
  &&-P(t)\hat{M}_x+K(t)\overline{\hat{M}}_z\hat{M}_y=0.
\end{eqnarray}
Here $\hat{\mathbf{M}}=\mathbf{M}/|\mathbf{M}|$ is the unit vector along
the free layer magnetization, $\overline{\hat{M}}_z=\pm1$ is the original
stationary collinear position of the vector, $\hat{M}_{x,\,y}$ are the
small deviations from that position, $\mathbf H$ is the external magnetic
field, $H_a$ is the anisotropy field of the free layer, $\kappa$ is the
Gilbert damping constant ($\kappa\ll1$ is assumed), $\gamma$ is the
gyromagnetic ratio. We assume $|H_z|<H_a$ that excludes possibility of
switching the junction by magnetic field without current.

The functions $K(t)$ and $P(t)$ describe the spin-polarized current effect
for two different interaction mechanisms between the conduction electrons
and magnetic lattice, namely, spin transfer torque~\cite{Slonczewski,Berger}
and appearance of a region in the free layer with
nonequilibrium electron spin polarization~\cite{Heide,Gulyaev3}; for
brevity, we name these mechanisms as torque and injection ones,
respectively. Note that the contribution of the injection mechanism in the
Eqs.~(\ref{2}) and~(\ref{3}) has the same form as the magnetic field, so
that one may say about some effective magnetic field $H_{\rm eff}=H_z+P(t)/\gamma$.
Note, however, that unlike the similar effect of
the parametric instability under longitudinal pumping by high-frequency
magnetic field~\cite{Gurevich}, the spin-polarized current effect, as
mentioned above,  is of local nature with characteristic scale of order of
the spin diffusion length $l\sim10\div100$ nm.

Further we consider the case of high injection when the following inequalities fulfill:
\begin{equation}\label{4}
  Z_2/Z_1,\,Z_2/Z_3\ll\lambda\ll1,
\end{equation}
where
\begin{equation}\label{5}
  Z_i=\frac{l_i}{\sigma_i\left(1-Q_i^2\right)}\quad(i=1,\,2,\,3),
\end{equation}
$l_i$ is the spin diffusion length for $i$th layer,
$Q_i=(\sigma_{i\uparrow}-\sigma_{i\downarrow})/\sigma_i$ is the conduction
spin polarization, $\sigma_{i\uparrow},\,\sigma_{i\downarrow}$ are the
partial conductivities of the spin-up and spin-down electrons in $i$th
layer, $\sigma_i=\sigma_{i\uparrow}+\sigma_{i\downarrow}$ is the total
conductivity of the layer (see~\cite {Epshtein} for detail). With these
assumptions, the functions $K(t)$ and $P(t)$ take the form
\begin{equation}\label{6}
  P(t)=\gamma H_a\frac{j(t)}{j_0}\equiv P_0+P_1\cos\Omega t,
\end{equation}
\begin{equation}\label{7}
  K(t)=\frac{P(t)}{\alpha\tau\gamma M}\equiv K_0+K_1\cos\Omega t,
\end{equation}
where $\alpha$ is the dimensionless constant of the \emph{sd} exchange interaction,
\begin{equation}\label{8}
  j_0=\frac{eH_aL}{\mu_B\alpha\tau Q_1},
\end{equation}
$\mu_B$ is the Bohr magneton, $j_0$ has the meaning of the threshold
direct current density, which corresponds to occurring instability of the
antiparallel configuration of the magnetic junction under dominating
injection mechanism without external magnetic field.

Let us make the following substitution in Eqs.~(\ref{2}),~(\ref{3}):
\begin{eqnarray}\label{9}
    \hat{M}_x(t)=X(t)\exp\left\{-\frac{K_1\overline{\hat{M}}_z}{\Omega}\sin\Omega
    t\right\},\nonumber \\
    \hat{M}_y(t)=Y(t)\exp\left\{-\frac{K_1\overline{\hat{M}}_z}{\Omega}\sin\Omega
    t\right\}.
\end{eqnarray}
The following equations are obtained for functions $X(t),\,Y(t)$:
\begin{eqnarray}\label{10}
  &&\frac{dX}{dt}+\kappa\overline{\hat{M}}_z\frac{dY}{dt}+\left[\gamma\left(H_z
  +H_a\overline{\hat{M}}_z\right)+P_0\right]Y\nonumber \\
&&+K_0\overline{\hat{M}}_zX=-(P_1-\kappa K_1)Y\cos\Omega t,
\end{eqnarray}
\begin{eqnarray}\label{11}
  &&\frac{dY}{dt}-\kappa\overline{\hat{M}}_z\frac{dX}{dt}-\left[\gamma\left(H_z
  +H_a\overline{\hat{M}}_z+4\pi M\overline{\hat{M}}_z \right)+P_0\right]X\nonumber \\
&&+K_0\overline{\hat{M}}_zY=+(P_1-\kappa K_1)X\cos\Omega t,
\end{eqnarray}

The exponent in Eq.~(\ref{9}) varies within finite limits from
$\exp\left\{-K_1\overline{\hat{M}}_z/\Omega\right\}$ to
$\exp\left\{+K_1\overline{\hat{M}}_z/\Omega\right\}$, so that stability
conditions of the zero solutions for $X(t),\,Y(t)$ coincide with those for
$\hat{M}_x,\,\hat{M}_y$ in the linear approximation used here.

It is seen from Eqs.~(\ref{10}),~(\ref{11}) that the torque mechanism
contribution to the effects related with the alternative component of the
current ($K_1$) appears with the small factor $\kappa$ and is absent when
the damping is neglected.

Let us make the Fourier transformation in Eqs.~(\ref{10}),~(\ref{11}). The
equations for the Fourier components with $\omega$ frequency take the form
\begin{eqnarray}\label{12}
  &&-i\omega X(\omega)+\left[\gamma\left(H_z+H_a\overline{\hat{M}}_z\right)
  +P_0-i\kappa\omega\right]Y(\omega)\nonumber \\
&&+K_0\overline{\hat{M}}_zX(\omega)=-\epsilon\left[Y(\omega+\Omega)+Y(\omega-\Omega)\right],
\end{eqnarray}
\begin{eqnarray}\label{13}
  &&-i\omega Y(\omega)+\left[\gamma\left(H_z+H_a\overline{\hat{M}}_z+4\pi M\overline{\hat{M}}_z\right)
  +P_0-i\kappa\omega\right]X(\omega)\nonumber \\
&&+K_0\overline{\hat{M}}_zY(\omega)=+\epsilon\left[X(\omega+\Omega)+X(\omega-\Omega)\right],
\end{eqnarray}
where
\begin{equation}\label{14}
  \epsilon=\frac{1}{2}(P_1-\kappa K_1).
\end{equation}

There are ``shifted'' Fourier components with $\omega\pm\Omega$
frequencies in the right-hand side of Eqs.~(\ref{12}),~(\ref{13}). If one
writes the equations similar to~(\ref{10}),~(\ref{11}) for such
components, then components with $\omega$ and $\omega\pm2\Omega$
frequencies appear in the right-hand side of the equations. Continuing
this process leads to an infinite set of equations coupled via
$\epsilon$ parameter of frequency dimension. We assume this frequency to
be low compared to the other characteristic frequencies of the problem,
namely, the modulation frequency $\Omega$ and the system natural frequency
defined below. Therefore, only the component with $\omega$ frequency may
be kept in the right-hand side of the equations for the $\omega\pm\Omega$
components, so that closed set of equations is obtained.

In absence of the alternative component of the current ($\epsilon=0$), the
frequency $\omega$ is determined by the zero determinant condition for the
homogeneous set of equations
\begin{equation}\label{15}
  \Delta(\omega)\equiv-(1+\kappa^2)\omega^2-2i\beta\omega+\omega_0^2=0,
\end{equation}
where
\begin{equation}\label{16}
  \omega_0^2=\left(\Omega_x+P_0\overline{\hat{M}}_z\right) \left(\Omega_y+P_0\overline{\hat{M}}_z\right)+K_0^2,
\end{equation}
\begin{equation}\label{17}
  \beta=\kappa\left[\frac{\Omega_x+\Omega_y}{2}+\left(P_0
+\frac{K_0}{\kappa}\right)\overline{\hat{M}}_z\right],
\end{equation}
$$\Omega_x=\gamma\left(H_z\overline{\hat{M}}_z+H_a+4\pi M\right),\quad
\Omega_y=\gamma\left(H_z\overline{\hat{M}}_z+H_a\right)$$ (usually
$\Omega_x\gg\Omega_y$).

The fluctuations with frequency determined by Eq.~(\ref{15}) become
unstable even if one of the following conditions fulfils:
\begin{equation}\label{18}
  \omega_0^2<0,\quad\beta<0.
\end{equation}

In presence of the torque mechanism only ($P_0=0$), only the second of
conditions (\ref{18}) can be fulfilled at
$K_0>\displaystyle\frac{\kappa}{2}(\Omega_x+\Omega_y)$,
$\overline{\hat{M}}_z=-1$ (the antiparallel configuration). In presence of the
injection mechanism only ($K_0=0$), the first condition fulfils at
$P_0>\Omega_y$, $\overline{\hat{M}}_z=-1$. By choosing the external magnetic
field close enough (in magnitude) to the anisotropy field ($0<H_a-H_z\ll
H_a$), the injection mechanism instability threshold can be lowered
significantly. Such a field does not influence practically the torque
mechanism threshold because of $\Omega_x\gg\Omega_y$. Under such
conditions the torque mechanism may be neglecting, so that $K_0=0$ may be laid in
Eqs.~(\ref{16}),~(\ref{17}).

We seek solution of the dispersion equation in presence of the
high-frequency component in the form
$\omega=\omega_0+\nu,\,|\nu|\ll|\omega_0|$. We are interesting below in
stability of the solution under fulfilling the parametric resonance
condition $\Omega=2\omega_0$.

By virtue of small $\epsilon$ parameter, we may neglect the terms in the
dispersion equation where that parameter is multiplied by small quantities
$\kappa$ and $\nu$.

Under actual conditions, $4\pi M\gg H_a,\,H,\,P_0$ inequalities
fulfill usually. With these inequalities taking into account, the
following parametric resonance equation is obtained:
\begin{equation}\label{19}
  \Delta(\omega)\Delta(\omega-\Omega)=(4\pi\gamma M\epsilon)^2.
\end{equation}

We obtain from Eq.~(\ref{19})
\begin{equation}\label{20}
  \omega=\omega_0-2\pi i\kappa\gamma M\pm\frac{2i\pi\gamma M\epsilon}{\omega_0},
\end{equation}
\begin{eqnarray}\label{21}
  &&\omega_0=\sqrt{\left(\Omega_x+P_0\overline{\hat{M}}_z\right)\left(\Omega_y
  +P_0\overline{\hat{M}}_z\right)}\nonumber \\
  &&\approx\sqrt{4\pi\gamma M\left[\gamma\left(H_z\overline{\hat{M}}_z
  +H_a\right) +P_0\overline{\hat{M}}_z\right]}.
\end{eqnarray}

The following condition of the parametric instability is obtained from Eq.~(\ref{20}):
\begin{equation}\label{22}
  \epsilon>\kappa\omega_0,\quad\omega_0^2>0.
\end{equation}

It follows from Eqs.~(\ref{18}) and~(\ref{22}) that two instability
mechanisms, namely, a nonparametric one (which is due to the spin
polarized direct current in absence of modulation) and parametric one due
to modulation, compete each other, so that only one of them is realized
for given configuration (parallel or antiparallel). However, a situation
is possible when instability of the antiparallel configuration is created
by the nonparametric mechanism, while the instability of the parallel
configuration is created by the parametric one. It is possible, also, that
both the collinear configurations become unstable due to the parametric
instability in absence of the nonparametric one. It is necessary to
compare the corresponding instability thresholds to know what of the
variants realizes.

The frequency $\omega_0$ takes different values for the parallel
($\overline{\hat{M}}_z=+1$) and antiparallel ($\overline{\hat{M}}_z=-1$) relative
orientation of the pinned and free layers, these values are denoted below
as $\omega_0^{(\pm)}$. At $|H_z|<H_a$ we have
$\left(\omega_0^{(+)}\right)^2>0$. It means stability of the parallel
configuration in absence of the current modulation and possibility of its
instability under high enough amplitude of the high-frequency component.

From Eq.~(\ref{21}) with Eqs.~(\ref{6}),~(\ref{8}) and~(\ref{14})
taking into account, we obtain the threshold value of the alternative
component amplitude $\widetilde j$ corresponding to occurring parametric
instability

\begin{equation}\label{23}
  \widetilde j_{\rm th}^{(\pm)}=\frac{2e\kappa L\omega_0^{(\pm)}}{\mu_B\gamma\alpha\tau Q_1}.
\end{equation}

At typical parameter values $\alpha\sim10^4$, $\tau\sim10^{-12}$ s,
$\kappa\sim10^{-2}$, $Q_1\sim10^{-1}$, $L\sim10^{-7}$ cm,
$\omega_0^{(\pm)}\sim10^{10}$ s$^{-1}$ we have $\widetilde j_{\rm
th}^{(\pm)}\sim10^4\div10^5$ A/cm$^2$.

The threshold direct current density
\begin{equation}\label{24}
  \overline j_{\rm th}^{(\pm)}=j_0(1\pm H_z/H_a)
\end{equation}
may be both larger and smaller than $\widetilde j_{\rm th}^{(\pm)}$; this
depends, in the first place, on the damping constant $\kappa$ and the
external magnetic field. Therefore, under the mentioned condition $\widetilde
j\le\overline j$ a situation is possible when the parametric instability
threshold is exceeded, while the nonparametric instability threshold still
does not reached.

Note that the threshold density of the high-frequency component $\widetilde
j_{\rm th}^{(-)}$ for the antiparallel configuration may be both smaller
and larger than similar quantity $\widetilde j_{\rm th}^{(+)}$ for the
parallel configuration; the latter situation takes place at $H_z<0$, $\overline
j/j_0<|H_z|/H_a$.

Let us consider the case $\widetilde j_{\rm th}^{(-)}<\widetilde j_{\rm
th}^{(+)}$. At $\overline j<\overline j_{\rm th}^{(-)}$, $\widetilde j_{\rm
th}^{(-)}<\widetilde j <\widetilde j_{\rm th}^{(+)}$ the parametric instability of
the antiparallel configuration takes place with stability of the parallel
one; in such a case the initial antiparallel configuration switches to
parallel one. At $\overline j<\overline j_{\rm th}^{(-)}$, $\widetilde j >\widetilde j_{\rm
th}^{(+)}$ parametric instability of both collinear configurations occurs,
so that an oscillation regime takes place. In the opposite case, $\widetilde
j_{\rm th}^{(-)}>\widetilde j_{\rm th}^{(+)}$, the parametric instability of
the parallel configuration takes place at $\overline j<\overline j_{\rm th}^{(-)}$,
$\widetilde j_{\rm th}^{(+)}<\widetilde j <\widetilde j_{\rm th}^{(-)}$ and both
collinear configurations at $\overline j<\overline j_{\rm th}^{(-)}$, $\widetilde j
>\widetilde j_{\rm th}^{(-)}$.

At $\overline j>\overline j_{\rm th}^{(-)}$ the nonparametric instability of the
antiparallel configuration occurs (because
$\left(\omega_0^{(-)}\right)^2<0$); in this case, as mentioned above,
parametric instability of this configuration is suppressed. However, at
$\widetilde j>\widetilde j_{\rm th}^{(\pm)}$ the parametric instability of the
parallel configuration takes place.

\begin{figure}
\includegraphics{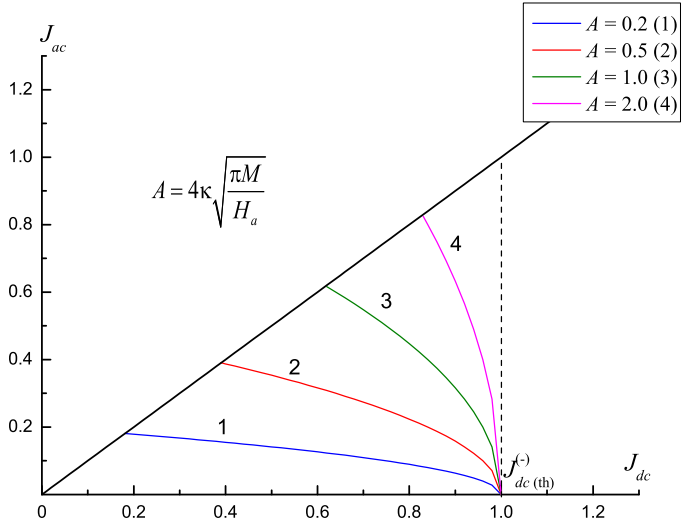}
\caption{Phase diagram for the antiparallel configuration (see explanation
in the text).}\label{fig2}
\end{figure}

In Fig.~\ref{fig2} a phase diagram in $\left(\overline j,\,\widetilde j\right)$
coordinates (with dimensionless variables $J_{dc}=\overline j/j_0,\,
J_{ac}=\widetilde j/j_0$) is shown for the antiparallel configuration at $H=0$
and various values of the parameter $A=4\kappa(\pi M/H_a)^{1/2}$. The
straight line bounds the range of the high-frequency current density
values according with the condition $\widetilde j\le\overline j$. On the right of
the dashed line $J_{dc}= J_{dc\,(\rm th)}^{(-)}\equiv\overline j_{\rm
th}^{(-)}/j_0$ the range is placed of the nonparametric instability of the
antiparallel configuration where $\left(\omega_0^{(-)}\right)^2<0$. The
parametric instability of this configuration at given value of the $A$
parameter takes place in the range between that line and the curve
corresponding to that value of $A$.

Thus, instability of any of two collinear configurations, as well as both
the configurations can be created under amplitude modulation of the
spin-polarized current. This opens possibility of the switching any of two
configurations to another (with the same direction of the total current),
as well creating an oscillatory regime when both collinear configurations
are unstable.

The experimentally observed lowering the threshold of switching by direct
current in presence of a high-frequency component may be related with
appearance of the parametric instability creating by that component under
direct current density lower then the threshold mentioned.

\section*{Acknowledgment}
The work was supported by the Russian Foundation for Basic Research,
Grants Nos.~08-07-00290-a and 10-02-00030-a.


\begin{thebibliography}{19}
\bibitem{Katine}
J.A. Katine, F.J. Albert, R.A. Buhrman, E.B. Myers, D.C. Ralph, Phys. Rev. Lett. \textbf{84}, 3149 (2000).
\bibitem{Tsoi}
M. Tsoi, A.G.M. Jansen, J. Bass, W.-C. Chiang, M. Seck, V. Tsoi, P. Wyder, Phys. Rev. Lett.
\textbf{80}, 4281 (1999).
\bibitem{Bass}
J. Bass, W.R. Pratt, Jr., J. Phys.: Condens. Matter \textbf{19}, 183201 (2007).
\bibitem{Krivorotov}
I.N. Krivorotov, N.C. Emley, J.C. Sankey, S.I. Kiselev, D.C. Ralph, R.A. Buhrman, Science
\textbf{307}, 228 (2005).
\bibitem{Ralph}
D.C. Ralph, M.D. Stiles, J. Magn. Magn. Mater. \textbf{320}, 1190 (2008).
\bibitem{Xiao}
Z.H. Xiao, X.Q. Ma, P.P. Wu, J.X. Zhang, L.Q. Chen, S.Q. Shi, J. Appl. Phys. \textbf{102}, 093907 (2007).
\bibitem{Rivkin}
K. Rivkin, J.B. Ketterson, Appl. Phys. Lett. \textbf{88}, 192515 (2006).
\bibitem{Cui}
Y.-T. Cui, J.C. Sankey, C. Wang, K.V. Thadani, Z.-P. Li, R.A. Buhrman, D.C. Ralph, Phys. Rev.
B \textbf{77}, 214440 (2008).
\bibitem{Florez1}
S.H. Florez, J.A. Katine, M. Carey, L. Folks, B.D. Terris, J. Appl. Phys. \textbf{103}, 07A708
(2008).
\bibitem{Florez2}
S.H. Florez, J.A. Katine, M. Carey, L. Folks, O. Ozatay, B.D. Terris, Phys. Rev.
B \textbf{78}, 184403 (2008).
\bibitem{Biziere}
N. Biziere, E. Mur\'e, J.-Ph. Ansermet, J. Magn. Magn. Mater. \textbf{322}, 3320 (2010).
\bibitem{Gulyaev1}
Yu.V. Gulyaev, P.E. Zilberman, A.I. Krikunov, A.I. Panas, E.M. Epshtein, JETP Lett. \textbf{86}
328 (2007).
\bibitem{Gulyaev2}
Yu.V. Gulyaev, P.E. Zilberman, A.I. Panas, E.M. Epshtein, J. Exp. Theor. Phys. \textbf{107}, 1027 (2008).
\bibitem{Slonczewski}
J.C. Slonczewski, J. Magn. Magn. Mater. \textbf{159}, L1 (1996).
\bibitem{Berger}
L. Berger. Phys. Rev. B \textbf{54}, 9353 (1996).
\bibitem{Heide}
C. Heide, P.E. Zilberman, R.J. Elliott. Phys. Rev. B \textbf{63}, 064424 (2001).
\bibitem{Gulyaev3}
Yu.V. Gulyaev, P.E. Zilberman, E.M. Epshtein, R.J. Elliott, JETP Lett. \textbf{76}, 155 (2002).
\bibitem{Gurevich}
A.G. Gurevich, G.A. Melkov. Magnetizsation Oscillations and Waves (CRC Press, Boca Raton, FL, 1996).
\bibitem{Epshtein}
E.M. Epshtein, Yu.V. Gulyaev, P.E. Zilberman, J. Magn. Magn. Mater. \textbf{312}, 200 (2007).

\end{thebibliography}
\end{document}